\documentclass[12pt]{article}
\pdfoutput=1

\usepackage{epsf}
\usepackage{amssymb}
\usepackage{graphicx}
\usepackage{amsmath,amscd}

\def\dj{\hbox{d\kern-0.347em \vrule width 0.3em height 1.252ex depth
-1.21ex \kern 0.051em}}

\numberwithin{equation}{section}

\begin{document}

\setlength{\oddsidemargin}{0cm}
\setlength{\baselineskip}{7mm}


\thispagestyle{empty}
\setcounter{page}{0}

\begin{flushright}

\end{flushright}

\vspace*{1cm}

\begin{center}
{\bf \Large Colliding AdS gravitational shock waves in various}

\vspace*{0.3cm}

{\bf \Large  dimensions and holography}

\vspace*{0.3cm}


\vspace*{1cm}

\'Alvaro Due\~nas-Vidal\footnote{\tt 
adv@usal.es}
and Miguel A. V\'azquez-Mozo\footnote{\tt 
Miguel.Vazquez-Mozo@cern.ch}

\end{center}

\vspace*{0.0cm}

\begin{center}
  
 {\sl Departamento de F\'{\i}sica Fundamental,
 Universidad de Salamanca \\ 
 Plaza de la Merced s/n,
 E-37008 Salamanca, Spain
  }

and

{\sl Instituto Universitario de F\'{\i}sica Fundamental y Matem\'aticas (IUFFyM) \\
Universidad de Salamanca, Salamanca, Spain}

\end{center}

\vspace*{1.5cm}

\centerline{\bf \large Abstract}

\noindent
The formation of marginally trapped surfaces in the off-center 
collision of two shock waves on AdS$_{D}$ (with $D=4,5,6,7$ and $8$)
is studied numerically. We focus on the case when the two waves collide with nonvanishing impact parameter while the sources
are located at the same value of the holographic coordinate. In all cases a critical value of the impact parameter is found
above which no trapped surface is formed. The numerical results show the existence of 
a simple scaling relation between the critical impact
parameter and the energy of the colliding waves.
Using the isometries of AdS$_{D}$ we relate the solutions obtained
to the ones describing the collision of two waves with a purely holographic impact parameter. This provides a 
gravitational dual for the head-on collision of two lumps of energy of unequal size.

\newpage  

\setcounter{footnote}{0}

\section{Introduction}

The applications of AdS/CFT duality 
to the description of heavy ion collisions has attracted quite a lot of attention in recent years,
in the hope of describing the phenomenology observed in RHIC and expected in the LHC 
(see \cite{review_hic_ads} for reviews).
The detection in RHIC of a large elliptic flow and jet quenching are interpreted as indications
that the quark-gluon plasma produced in the collision is strongly coupled, thus making AdS/CFT a very useful
tool to capture some of the relevant physics involved. 
In spite of the difference between real QCD and $\mathcal{N}=4$ SYM,
the calculations of some relevant observables using gravitational duals
\cite{trials} show a reasonable good agreement with the experimental results.

A possible way to model the collision of two heavy ions in a strongly coupled gauge theory is
to consider the scattering of two energy lumps in a $\mathcal{N}=4$ SYM theory at strong coupling \cite{janik_peschanski,gpy1}. The gravitational dual of such a system is a
space-time representing the collision of two shock gravitational waves propagating in AdS$_{5}$.
Nonetheless, the picture of the heavy ions proposed in \cite{gpy1} is not free of problems \cite{perils}. 
For example, it produces an energy density profile for the lumps 
in the gauge theory that falls off
as a power of the transverse distance from its center, quite unlike the exponential fall off expected from phenomenological 
nuclear potentials. However, in spite of these caveats, these systems are a good test bench to analyze  
collective properties of strongly coupled plasmas that might be of relevance in understanding the phenomenology of heavy ion 
collisions \cite{collective}. 

In this paper we study various aspects of the collision of shock gravitational waves in AdS space-time in various dimensions.
The metric in the region before the collision takes place (i.e., where at least one of the 
two null coordinates $u$ or $v$ is negative as shown in Fig. \ref{fig:1}) is given by
\begin{eqnarray}
ds^{2}={L^{2}\over z^{2}}\left[dz^{2}-dudv+d\vec{x}^{\,2}
+{z\over L}\Phi_{+}(z,\vec{x})\delta(u)du^{2}+
{z\over L}\Phi_{-}(z,\vec{x})\delta(v)dv^{2}\right],
\label{metric}
\end{eqnarray}
where $\Phi_{\pm}(z,\vec{x})$ are the wave profiles and 
$L$ is the AdS$_{D}$ radius. A full description of the final fate of the wave collision (and its holographic 
interpretation) would require knowing in detail
the solution to the Einstein equations in the interaction region $u>0$, $v>0$, something
that up to date has not been fully achieved (for some important progress in this direction see \cite{interaction}). 
Based on the intuition in flat space, however, one expects that the collision of two
shock waves would produce some kind of dressed singularity. The holographic interpretation of
such a process would be the thermalization of the $\mathcal{N}=4$ SYM plasma produced as the result  
of the collision of the two incoming lumps whose energy distribution is given by a CFT energy momentum
tensor with components \cite{gpy2}
\begin{eqnarray}
\langle T_{uu}\rangle_{\rm CFT}&=&{2^{D-2}\mu_{+}\over {\rm Vol}(S^{D-3})}{z_{+}^{D-1}\over [(\vec{x}-\vec{b}_{+})^{2}+z_{+}^{2}]^{D-2}}
\delta(u), \nonumber \\
\langle T_{vv}\rangle_{\rm CFT}&=&{2^{D-2}\mu_{-}\over {\rm Vol}(S^{D-3})}{z_{-}^{D-1}\over [(\vec{x}-\vec{b}_{-})^{2}+z_{-}^{2}]^{D-2}}
\delta(v),
\label{dualTuuTvv}
\end{eqnarray}
with $(z_{\pm},\vec{b}_{\pm})$ the holographic and transverse coordinates 
of the sources of the waves \eqref{metric}.

A way to avoid solving for the geometry in the $u>0$, $v>0$ wedge is to look for 
the formation of an apparent horizon before actually getting into the region of space-time where the 
interaction takes place \cite{flat}. In AdS for the case of zero impact
parameter it was found in \cite{gpy1} that such a trapped surface is always formed for any energy of the incoming waves.
The case of collisions with a nonvanishing impact parameter along the field theory coordinates has been studied in AdS$_{5}$ \cite{lin_shuryak}, where there is an energy-dependent critical value of the
impact parameter above which no marginally trapped surface of the type sought is formed. The results found
in \cite{gpy1} and \cite{lin_shuryak} are qualitatively similar to the corresponding situations in flat spacetime 
\cite{flatip,flatip2}.  

In the dual $\mathcal{N}=4$ SYM theory the interpretation of the results of \cite{lin_shuryak} would be the 
existence of a critical impact parameter for the thermalization of the plasma following the collision
of two energy lumps of the same size. This can be understood if we think that for large enough impact parameter
the two energy distributions do not have enough overlap to induce a thermalization of the whole plasma once the collision 
takes places. Once again, to see what happens as the result of the collision would require solving the 
field equations into the interaction region\footnote{We should not forget that the non-existence of the trapped surface
of Penrose's type \cite{flat} does not exclude that other trapped surfaces are formed in the interaction region.}. 

Another interesting physical situation is that of the collision of two shock waves whose sources have different values of the
holographic coordinates. A look at Eq. \eqref{dualTuuTvv} shows that in the holographic theory this describes the collision
of two energy lumps of different size, with or without impact parameter. This problem has been addressed in 
\cite{gpy2}, where the trapped surface equation is solved analytically 
in the limit in which the impact parameter is much smaller than the critical value. 
It would be interesting, nevertheless, to study the collision of waves with ``large'' holographic impact parameter 
in various dimensions.

In Ref. \cite{us} the formation of trapped surfaces in the head-on collision of two shock waves was studied
for incoming waves characterized by a finite size in transverse space, in contrast with the case studied in \cite{gpy1} where the 
energy density along the wave-front follows a delta-function distribution in transverse coordinates. 
For $D=4$ and $D=5$ it was found that the marginally closed trapped surface only forms if the transverse size of the wave is 
smaller than certain critical value.  If we interpret the formation of the 
trapped surface as a signal of eventual black hole formation as a result of the collision, this could imply at face value 
the existence of a threshold for thermalization in the holographic gauge theory as a function of the spread of the wave
in the gravitational dual. 
It is not clear, however, what meaning the spread of the gravitational sources introduced in \cite{us} might have
in the holographic conformal field theory. For example, a calculation of the holographic energy-momentum tensor shows that the 
size of the energy distribution in the boundary theory is independent of the value of the deformation parameter.
Indeed, applying the holographic prescription \cite{dhss} to the
solution found in \cite{us} for the wave profile $\Phi(z,\vec{x})$, the associated 
holographic energy-momentum tensor is 
\begin{eqnarray}
\langle T_{uu}\rangle_{\rm CFT}={2^{D-2}\mu\over {\rm Vol\,}(S^{D-3})}{L^{D-1}\over (\vec{x}^{\,2}+L^{2})^{D-2}}\delta(u),
\end{eqnarray} 
and similarly for $\langle T_{vv}\rangle_{\rm CFT}$ by replacing $u\rightarrow v$. Thus no trace of the smearing of the gravitational
source is left in the energy distribution in the boundary theory, 
its transverse size being determined solely by the value of the holographic coordinate 
of the source, $z=L$.

In this note our aim is twofold: first to study numerically the formation of marginally closed trapped 
surfaces in the collision of two shock waves in 
AdS$_{D}$ space-time with a nonvanishing impact parameter in the field theory coordinates, thus 
extending the analysis of  \cite{lin_shuryak} to $D\neq 5$. Second, to apply the results obtained to 
the analysis of collisions of
two AdS$_{D}$ shock waves with ``purely holographic'' impact parameter, i.e. when the sources of the two incoming waves 
are located at points in transverse space that only differ in the value of the holographic coordinate. 
The strategy used consists in exploiting the underlying O(2,$D-1$) isometries of 
AdS$_{D}$  
to map this problem to a problem of the scattering of two waves with ``spatial" impact parameter
(cf. \cite{gpy2,lin_shuryak}).  
As explained above, in the spirit of the AdS/CFT correspondence, collisions with pure holographic impact parameter
provide the gravitational 
dual of the high-energy 
head-on collision of two energy distributions of unequal size in the strongly coupled CFT.

In Section \ref{sec:gw} we review the series of isometries of AdS$_{D}$ that allow to rotate a general collision with
purely holographic impact parameter into a symmetric collision with an impact parameter along the field theory coordinates. 
The numerical analysis
of the trapped surface equations is done in Section \ref{sec:cts}. Finally, our results 
together with their holographic interpretation are discussed in Section \ref{sec:concl}. To make the presentation 
self-contained some technical details have been deferred to the
Appendix.

\section{Gravitational shock wave collisions with holographic impact parameter}
\label{sec:gw}

\begin{figure}
\centerline{\includegraphics[width=3.0in]{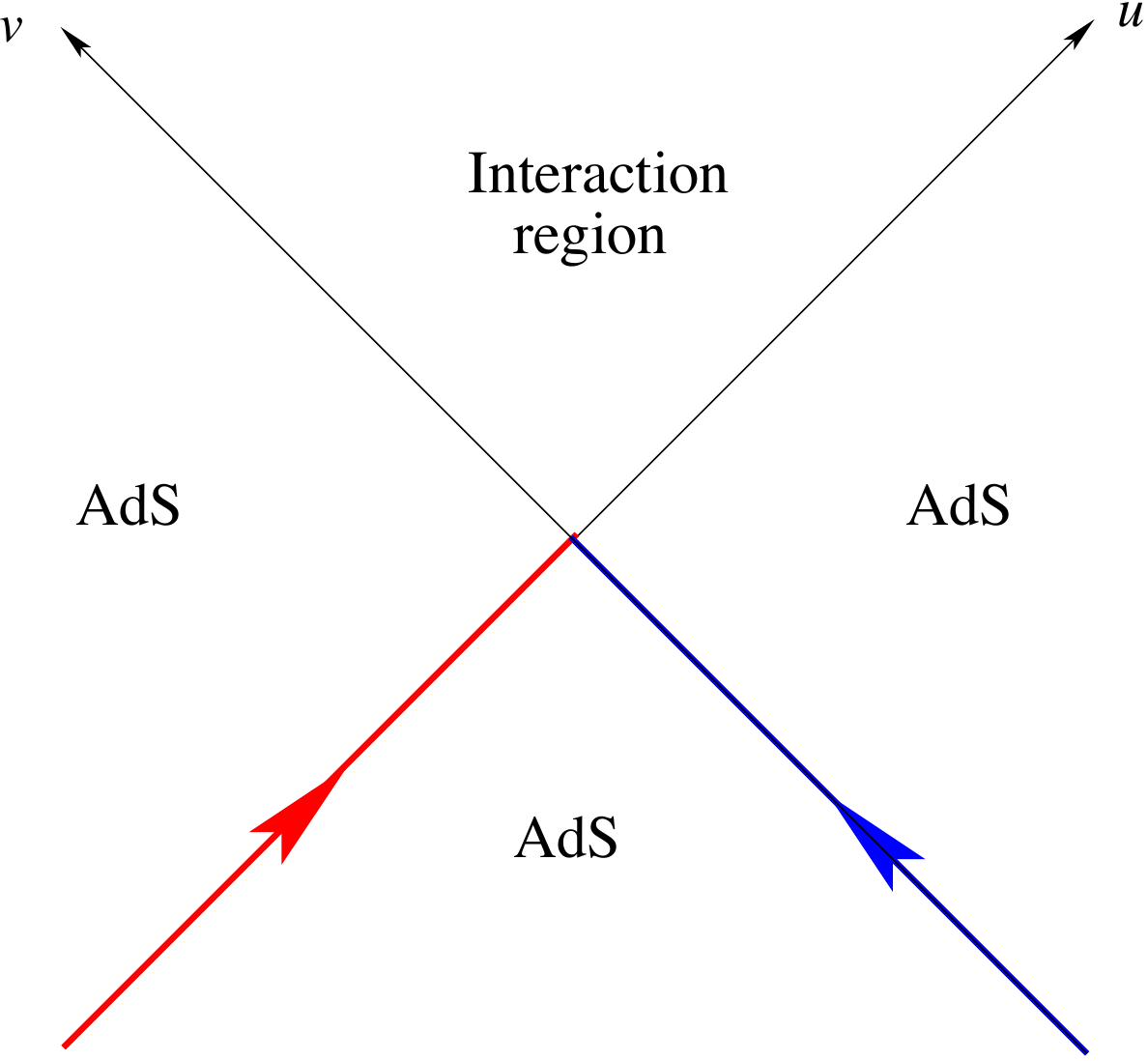}}
\caption{The four regions in the space-time describing the interaction of two shock waves in AdS$_{D}$. The blue (right)
and red (left) thick lines represent the location of the incoming waves, supported respectively at $u=0$, $v<0$ and $u<0$, $v=0$.}
\label{fig:1}
\end{figure}

We study the collision of two shock waves propagating on AdS$_{D}$ whose delta-function sources are
located in the hyperbolic transverse space $\mathbb{H}_{D-2}$ at the points $(z_{\pm},\vec{b}_{\pm})$. 
This space can be seen as a two-sheeted hyperboloid 
\begin{eqnarray}
(Y^{0})^{2}-Y^{(D-2)}-\sum_{i=1}^{D-3}(Y^{i})^{2}=L^{2}
\end{eqnarray}
in $(D-1)$-dimensional Minkowski space. Poincar\'e coordinates $(z,\vec{x})$ can be introduced 
\begin{eqnarray}
Y^{0}&=& {z\over 2}\left(1+{L^{2}+\vec{x}^{\,2}\over z^{2}}\right), \nonumber \\
Y^{a}&=& {L\over z}x^{a}, \hspace*{4cm} a=1,\ldots,D-3, \\
Y^{D-2}&=& {z\over 2}\left(-1+{L^{2}-\vec{x}^{\,2}\over z^{2}}\right), \nonumber
\end{eqnarray}
in which the metric reads
\begin{eqnarray}
ds^{2}_{\mathbb{H}_{D-2}}={L^{2}\over z^{2}}\Big(dz^{2}+d\vec{x}^{\,2}\Big).
\end{eqnarray}

The metric of the full space-time previous to the collision 
is of the form shown in Eq. \eqref{metric} and
is characterized by the profile functions $\Phi_{\pm}(z,\vec{x})$. The Einstein equations reduce, outside the 
interaction region in Fig. \ref{fig:1}, to the Poisson-like equation in transverse space
\begin{eqnarray}
\left(\Box_{\mathbb{H}_{D-2}}-{D-2\over L^{2}}\right)\Phi_{\pm}(z,\vec{x})=
-16\pi G_{N}\mu_{\pm}\left({z_{\pm}\over L}\right)^{D-1}\delta(z-z_{\pm})\delta^{(D-3)}(\vec{x}-\vec{b}_{\pm}).
\label{ee}
\end{eqnarray}
Here $\Box_{\mathbb{H}_{D-2}}$ is the Laplacian in $\mathbb{H}_{D-2}$ 
\begin{eqnarray}
\Box_{\mathbb{H}_{D-2}}={z^{D-2}\over L^{2}}\partial_{z}\Big(z^{4-D}\partial_{z}\,\,\,\,\Big)
+{z^{2}\over L^{2}}\vec{\nabla}^{2}
\end{eqnarray}
and 
$\mu_{\pm}$ is the energy of the incoming waves, as read from the energy-momentum tensor
\begin{eqnarray}
T_{uu}&=&\mu_{+}\left({z_{+}\over L}\right)^{D-2}\delta(u)\delta(z-z_{+})\delta^{(D-3)}(\vec{x}-\vec{b}_{+}), \nonumber \\
T_{vv}&=&\mu_{-}\left({z_{-}\over L}\right)^{D-2}\delta(v)\delta(z-z_{-})\delta^{(D-3)}(\vec{x}-\vec{b}_{-}).
\label{emtensors}
\end{eqnarray}
In what follows we focus on the family of incoming trajectories considered in \cite{gpy1} where the 
sources are located at the points $(z_{\pm},b_{\pm},0,\ldots,0)$ with
\begin{eqnarray}
z_{\pm}(a)&=&{L\over {\sqrt{1+\beta^{2}}\pm\beta\cos{a}}}, \nonumber \\
b_{\pm}(a)&=& \pm {L\beta\sin{a}\over \sqrt{1+\beta^{2}}\pm\beta\cos{a}},
\label{sources_points}
\end{eqnarray}
with $0\leq a<2\pi$. 

The space $\mathbb{H}_{D-2}$ has an O(1,$D-2$) group of isometries. This can be seen as the part of the  
full O(2,$D-1$) isometry of AdS$_{D}$ that preserves the null coordinates $u,v$.  
The group O(1,$D-2$) is generated by the O($D-2$) rotations of the 
ambient space coordinates $(Y^{1},\ldots,Y^{D-2})$
among themselves, plus boost along each of these ``spatial" coordinates. 
In the following we describe a series of isometries of AdS$_{D}$ (and therefore of $\mathbb{H}_{D-2}$) 
that connect a general collision of two waves with sources at $({L\over \sqrt{1+\beta^{2}}\pm\beta},0,0,\ldots)$ to 
a symmetric collision where the sources are located at $(L,\pm L\beta,0,\ldots)$. 

\paragraph{O(2) rotations.}

To begin with we are going to focus on the O(2) subgroup of O($D-2$) that rotates
the embedding coordinates $Y^{1}$, $Y^{D-2}$ 
\begin{eqnarray}
Y'{}^{1}&=& Y^{1}\cos{\theta}+Y^{D-2}\sin{\theta}, \nonumber \\
Y'{}^{D-2}&=& -Y^{1}\sin{\theta}+Y^{D-2}\cos{\theta}.
\end{eqnarray}
In terms of the Poincar\'e coordinates $(z,\vec{x})=
(z,x,\vec{x}_{T})$ the O(2) rotations take the form
\begin{eqnarray}
z'&=& {2L^{2}z\over L^{2}+z^{2}+x^{2}+\vec{x}^{\,2}_{T}+(L^{2}-z^{2}-x^{2}
-\vec{x}^{\,2}_{T})\cos{\theta}-2Lx\sin{\theta}}, \nonumber \\[0.3cm]
x'&=& {2L^{2}x\cos{\theta}+L(L^{2}-z^{2}-\vec{x}^{\,2}_{\perp})\sin{\theta}\over
L^{2}+z^{2}+x^{2}+\vec{x}^{\,2}_{T}+(L^{2}-z^{2}-x^{2}
-\vec{x}^{\,2}_{T})\cos{\theta}-2Lx\sin{\theta}},  
\label{fullrotations}\\[0.3cm]
\vec{x}_{T}{}\!\!'&=& {2L^{2} \vec{x}_{T}\over L^{2}+z^{2}+x^{2}+\vec{x}^{\,2}_{T}+(L^{2}-z^{2}-x^{2}
-\vec{x}^{\,2}_{T})\cos{\theta}-2Lx\sin{\theta}}. \nonumber
\end{eqnarray}
In spite of the nonlinear action on the coordinates it can be checked without much effort that the previous
transformations form an abelian group. Actually, the O(2) rotations of the coordinates of $\mathbb{H}_{D-2}$ 
can be written as the one-parameter group of transformations
\begin{eqnarray}
\mathcal{U}(\theta)=\exp\left[i{\theta\over 2}\left({1\over L}\mathcal{K}_{x}+L\mathcal{P}_{x}\right)\right],
\label{finite_transf_bulk}
\end{eqnarray}
where 
\begin{eqnarray}
\mathcal{P}_{j}&=&  -i\partial_{j} \nonumber \\
\mathcal{K}_{j}&=& i\Big(z^{2}+x^{2}+\vec{x}_{T}^{\,2}\Big)\partial_{j}-2ix_{j}
\Big(z\partial_{z}+x\partial_{x}+\vec{x}_{T}\cdot\vec{\nabla}_{T}\Big)
\end{eqnarray}
are respectively the generators of translations and ``special conformal transformations'' on the
field theory coordinates $(x,\vec{x}_{T})$. This transformation can be lifted to the corresponding 
isometry of the full AdS$_{D}$ space-time.

Acting on the family of points shown in Eq. 
(\ref{sources_points}) the O(2) rotations amount to a translation in the
parameter $a$
\begin{eqnarray}
z_{\pm}'(\theta)=z_{\pm}(a+\theta), \hspace*{1cm}
b_{\pm}'(\theta)=b_{\pm}(a+\theta).
\end{eqnarray}
Hence, by performing an isometry of the background $\mathbb{H}_{D-2}$ we can transform the points 
\begin{eqnarray}
z_{\pm}={L\over \sqrt{1+\beta^{2}}\pm \beta}, \hspace*{1cm}
b_{\pm}=0, 
\label{sources1}
\end{eqnarray}
separated in the holographic direction $z$, into 
\begin{eqnarray}
z_{\pm}'(\theta)&=&{L\over {\sqrt{1+\beta^{2}}\pm\beta\cos{\theta}}}, \nonumber \\
b_{\pm}'(\theta)&=& \pm {L\beta\sin{\theta}\over \sqrt{1+\beta^{2}}\pm\beta\cos{\theta}}.
\label{sources2}
\end{eqnarray}
This O(2) transformation maps a collision of two shock waves with 
holographic impact parameter into a problem in which the two waves collide generically with both 
spatial and holographic impact parameter.

In our original problem \eqref{ee} the group of isommetries O(1,$D-2$), and therefore its O(2) subgroup 
\eqref{fullrotations}, 
is broken only by the presence of the sources. Because of this one can use 
this underlying symmetry as a solution generating technique, since the transformation of a solution to
the Einstein equations by an isometry is again a solution where the position of the sources is also tranformed.
To see this we first compute the Jacobian associated with \eqref{fullrotations}  
\begin{eqnarray}
\left|{\partial(z',x',\vec{x}_{T}\,')\over \partial(z,x,\vec{x}_{T})}\right|&=&
\left[{2L^{2}\over L^{2}+z^{2}+x^{2}+\vec{x}^{\,2}_{T}+(L^{2}-z^{2}
-\vec{x}^{\,2}_{\perp})\cos{\theta}-2Lx^{1}\sin{\theta}}\right]^{D-2} \nonumber \\[0.3cm]
&=& \left({z'\over z}\right)^{D-2}.
\end{eqnarray}
Now we apply \eqref{fullrotations} to the equations \eqref{ee} with sources located at \eqref{sources1}.
The Laplacian is invariant, $\Box_{\mathbb{H}_{D-2}}=\Box_{\mathbb{H}_{D-2}}'$, 
since we are dealing with an isometry of $\mathbb{H}_{D-2}$. On the other hand, the 
transformation of the delta functions can be computed using the expression of the Jacobian, leading to the 
result
\begin{eqnarray}
\left(\Box_{\mathbb{H}_{D-2}}'-{D-2\over L^{2}}\right)\Phi_{\pm}(z',\vec{x}\,')=\hspace*{8cm}
\nonumber \\
-16\pi G_{N}\mu_{\pm}'(\theta)\left[{z_{\pm}'(\theta)\over L}\right]^{D-1}\delta\Big(z'-z_{\pm}'(\theta)\Big)
\delta\Big({x}'-{b}_{\pm}'(\theta)\Big)\delta^{(D-4)}(\vec{x}_{T}\,\!\!').
\label{eqtrans1}
\end{eqnarray}
This describes the collision of two AdS$_{D}$ waves with sources at \eqref{sources2}
and energies
\begin{eqnarray}
\mu_{\pm}'(\theta)=\mu_{\pm}\left({\sqrt{1+\beta^{2}}\pm\beta\cos{\theta}\over\sqrt{1+\beta^{2}}\pm\beta}
\right).
\label{energy_after_rotation}
\end{eqnarray}

\begin{figure}[t]
\centerline{\includegraphics[width=.8\columnwidth]{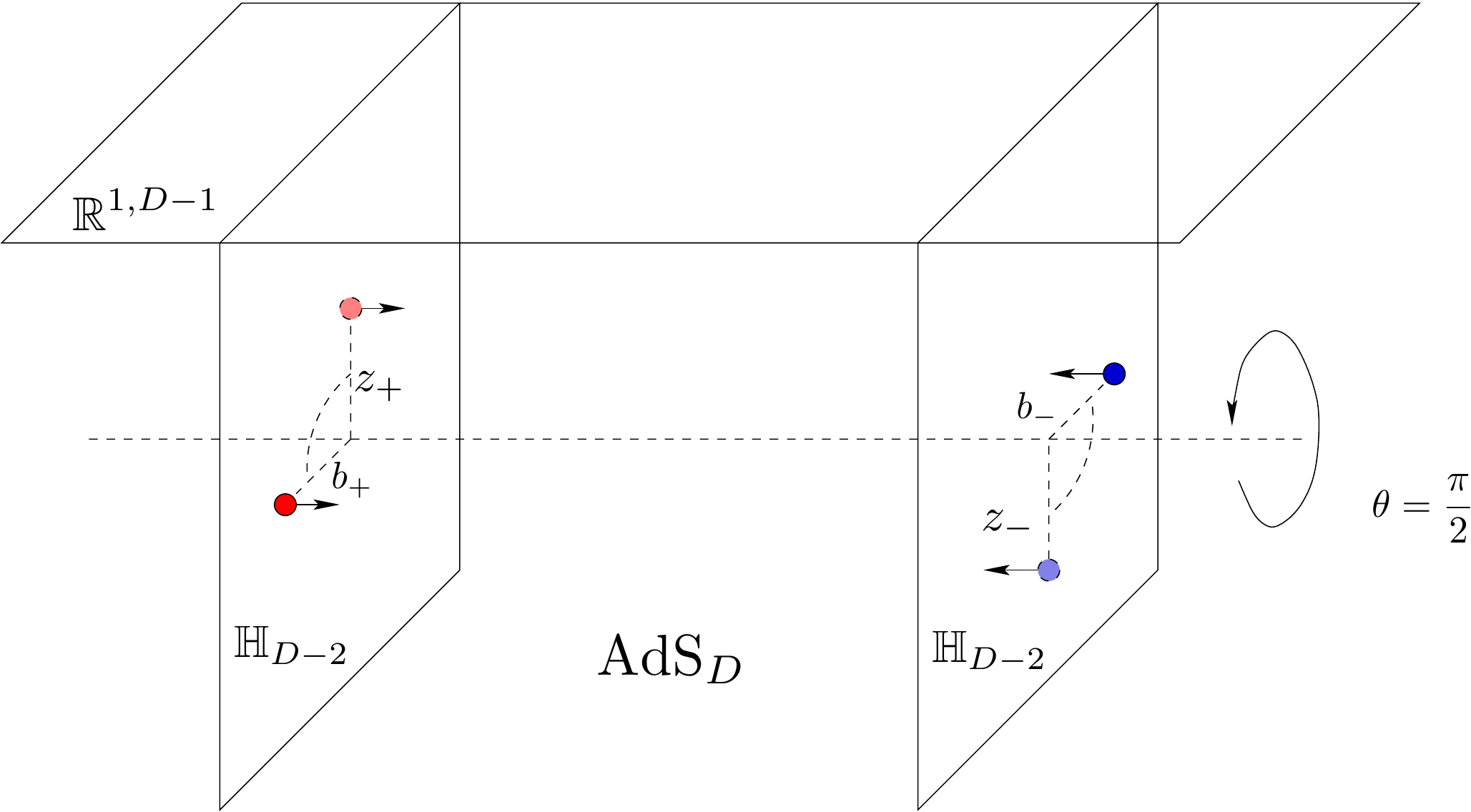}}
    \caption{Schematic representation of the action of the O(2) rotation on the sources of the
    two colliding waves.}
  \label{fig:2}
\end{figure}

This transformation of the energies 
$\mu_{\pm}$ as the result of this rotation can be understood on physical grounds
remembering that they correspond to the energy of the incoming field theory energy lumps at the boundary. 
The O(2) transformation on the transverse hyperbolic space $\mathbb{H}_{D-2}$ brings one of the sources 
closer to the AdS boundary whereas the other ``moves'' deeper into the bulk (see Fig. \ref{fig:2}). 
From the point of view of the
CFT on the boundary this means that the energy of the configuration associated with the first source 
increases whereas the energy of the lump associated with the second source gets smaller. This is a simple 
consequence of the IR/UV connection in AdS/CFT. 

We now particularize Eq. \eqref{eqtrans1} to the case $\theta={\pi\over 2}$, where the $z$ coordinate
of the two sources are equal. We find that the Einstein equations read in this case
\begin{eqnarray}
\left(\Box_{\mathbb{H}_{D-2}}'-{D-2\over L^{2}}\right)\Phi_{\pm}(z',\vec{x}\,')=\hspace*{8cm}
\nonumber \\
-16\pi G_{N}\mu_{\pm}'\left({z_{\pm}'\over L}\right)^{D-1}\delta\Big(z'-z_{\pm}'\Big)
\delta\Big({x}'-{b}_{\pm}'\Big)\delta^{(D-4)}(\vec{x}_{T}\,\!\!')
\label{eq_step1}
\end{eqnarray}
and describes now the problem of the collision of two waves with energies 
\begin{eqnarray}
\mu_{\pm}'=\mu_{\pm}{\sqrt{1+\beta^{2}}\over \sqrt{1+\beta^{2}}\pm\beta},
\end{eqnarray}
whose sources are located at
\begin{eqnarray}
z'_{\pm}={L\over \sqrt{1+\beta^{2}}}, \hspace*{1cm} b_{\pm}'=\pm {L\beta\over\sqrt{1+\beta^{2}}}.
\label{sources_step1}
\end{eqnarray}
This means that we have transformed our problem into the collision of two waves with spatial impact 
parameter.

\paragraph{Longitudinal boosts.}

This is however not the whole story. 
We have started with a collision where the two waves had different energy and, in general, this is also 
the case after the O(2) rotation has been performed. For practical purposes, however, it is 
simpler to study symmetric collisions where the two waves approach each other at the same energy. To achieve
this we can use another isometry of AdS$_{D}$ consisting in longitudinal boosts on the $u$, $v$ coordinates. 
To see how this affects the shock wave we apply the boost
\begin{eqnarray}
\overline{u}=\lambda u,\hspace*{1cm} \overline{v}=\lambda^{-1} v, \hspace*{1cm} \lambda>0
\end{eqnarray}
to the shock wave metric \eqref{metric}. The metric preserves the same form now with the profile
\begin{eqnarray}
\overline{\Phi}_{\pm}(z',\vec{x}\,')= \lambda^{\mp 1}\Phi_{\pm}(z',\vec{x}\,').
\label{transformation_profile}
\end{eqnarray}
Applying now the boost to the components of the energy-momentum tensor \eqref{emtensors} we find that
the energy $\mu_{\pm}$ associated with the transformed wave is
\begin{eqnarray}
\overline{\mu}_{\pm}'= \lambda^{\mp 1}\mu_{\pm}'.
\end{eqnarray}
Thus, by applying to Eq. \eqref{eq_step1} a longitudinal boost with parameter
\begin{eqnarray}
\lambda={\mu_{+}\over\mu_{-}}\left({\sqrt{1+\beta^{2}}-\beta\over\sqrt{1+\beta^{2}}+\beta}\right)
\label{longitudinal_boost}
\end{eqnarray}
we end up with a symmetric collision with energy $\sqrt{\mu_{+}\mu_{-}}$
\begin{eqnarray}
\left(\Box_{\mathbb{H}_{D-2}}'-{D-2\over L^{2}}\right)\overline{\Phi}_{\pm}(z',\vec{x}\,')=\hspace*{8cm}
\nonumber \\
-16\pi G_{N}\sqrt{\mu_{+}\mu_{-}}\left({z_{\pm}'\over L}\right)^{D-1}\delta\Big(z'-z_{\pm}'\Big)
\delta\Big({x}'-{b}_{\pm}'\Big)\delta^{(D-4)}(\vec{x}_{T}\,\!\!').
\label{eq_step2}  
\end{eqnarray}

\paragraph{Coordinate rescalings.}

There is however something unpleasant about the expressions of $z_{\pm}'$ and $b_{\pm}'$ shown in 
Eq. \eqref{sources_step1}. Since our only free parameter is $\beta$, 
we notice that it is not possible to change the spatial impact parameter $b_{\pm}'$
without at the same time 
moving the sources in the holographic direction $z_{\pm}$. To consider collisions where the spatial impact parameter
is decoupled from the value of the holographic coordinate of the sources we apply the following
rescaling to the coordinates
\begin{eqnarray}
z''=\sqrt{1+\beta^{2}}z', \hspace*{1cm} x''=\sqrt{1+\beta^{2}}x', \hspace*{1cm}
\vec{x}_{T}\,\!\!\!''=\sqrt{1+\beta^{2}}\vec{x}_{T}\,\!\!\!'.
\label{coordinates''}
\end{eqnarray}
In the full AdS$_{D}$ space this coordinate transformation is accompanied by similar rescaling of the two 
null coordinates. In the new coordinates the sources \eqref{sources_step1} are located in the points
\begin{eqnarray}
z''_{\pm}=L, \hspace*{1cm} b''_{\pm}=\mp L\beta,
\end{eqnarray}
thus decoupling the value of the holographic coordinate from that of 
the impact parameter. We have to keep in mind that
this rescaling is an isometry of $\mathbb{H}_{D-2}$.
Implementing it in Eq. \eqref{eq_step2} we arrive at
\begin{eqnarray}
\left(\Box_{\mathbb{H}_{D-2}}''-{D-2\over L^{2}}\right)\overline{\Phi}_{\pm}(z'',\vec{x}\,'')=\hspace*{8cm}
\nonumber \\
-16\pi G_{N}\sqrt{\mu_{+}\mu_{-}\over 1+\beta^{2}}\delta\Big(z''-L\Big)
\delta\Big({x}''\pm L\beta\Big)\delta^{(D-4)}(\vec{x}_{T}\,\!\!'').
\label{finaleq}
\end{eqnarray}
Now the two shock waves have spatial impact parameter $b\equiv |b_{+}''-b_{-}''|=2L\beta$ while both 
sources lie at $z''=L$. 

To summarize, we started with the collision of two shock waves with energies $\mu_{\pm}$ 
whose sources were located at the points $(z_{\pm},x_{\pm})$ given in Eq. 
\eqref{sources_step1}. 
Taking advantage of the isometries of AdS$_{D}$ (and of its transverse
section $\mathbb{H}_{D-2}$) this problem has been mapped into the collision described by Eq. \eqref{finaleq}
where the two sources are located at $z=L$ with an impact parameter $b$ given by
\begin{eqnarray}
b\equiv |b_{+}''-b_{-}''|=2L\beta=|z_{+}-z_{-}|\equiv \Delta z,
\end{eqnarray}
with $\Delta z$ the holographic impact parameter of the original collision. The change of coordinates
connecting Eqs. \eqref{eq_step1} and \eqref{finaleq} is given by 
\begin{eqnarray}
z''&=& {2L^{2}\sqrt{1+\beta^{2}}z\over L^{2}+z^{2}+x^{2}+\vec{x}^{\,2}_{T}+(L^{2}-z^{2}-x^{2}
-\vec{x}^{\,2}_{T})\cos{\theta}-2Lx\sin{\theta}}, \nonumber \\[0.3cm]
x''&=& {2L^{2}\sqrt{1+\beta^{2}}x\cos{\theta}+L\sqrt{1+\beta^{2}}
(L^{2}-z^{2}-\vec{x}^{\,2}_{\perp})\sin{\theta}\over
L^{2}+z^{2}+x^{2}+\vec{x}^{\,2}_{T}+(L^{2}-z^{2}-x^{2}
-\vec{x}^{\,2}_{T})\cos{\theta}-2Lx\sin{\theta}},  
\label{fullrotations_final}\\[0.3cm]
\vec{x}_{T}{}\!\!''&=& {2L^{2}\sqrt{1+\beta^{2}} \vec{x}_{T}\over L^{2}+z^{2}+x^{2}+\vec{x}^{\,2}_{T}+(L^{2}-z^{2}-x^{2}
-\vec{x}^{\,2}_{T})\cos{\theta}-2Lx\sin{\theta}}. \nonumber
\end{eqnarray}
To find the relation between the wave profiles we recall that these functions are only transformed by
the longitudinal boost\footnote{We have to keep in mind that in AdS$_{D}$ the rescaling \eqref{coordinates''}
also affects the null coordinates.}. Using the value \eqref{longitudinal_boost} for the parameter of this boost we
find that the solution $\Phi_{\pm}(z,\vec{x})$ for the collision with holographic impact parameter can 
be written in terms of the solution $\overline{\Phi}_{\pm}(z'',\vec{x}\,'')$ to Eq. \eqref{finaleq} as
\begin{eqnarray}
\Phi_{\pm}(z,\vec{x})={\mu_{-}\over\mu_{+}}\left({\sqrt{1+\beta^{2}}+\beta\over\sqrt{1+\beta^{2}}-\beta}
\right)^{\pm 1}\overline{\Phi}_{\pm}\Big(z''(z,\vec{x}),\vec{x}\,''(z,\vec{x})
\Big),
\end{eqnarray}
where the coordinate transformations $z''(z,\vec{x})$ and $\vec{x}\,''(z,\vec{x})$ are given by \eqref{fullrotations_final}.

The bottom line of this whole discussion is that the solution to the problem of finding a closed trapped
surface in the collision of two shock waves with an impact parameter aligned along the holographic direction 
is reduced to the one of finding this trapped surface in the collision of two shock waves with an impact
parameter that is purely spatial. This latter problem is easier to tackle numerically. Once the trapped surface
is found in this case the original one is obtained by applying the transformation \eqref{fullrotations_final}.
Because of the family of incoming trajectories \eqref{sources_points} used our discussion applies to any 
collision problem where the position of the sources in the holographic coordinate $z_{\pm}$ satisfies
$z_{+}z_{-}=L^{2}$. Notice, however, that for any other value of the product $z_{+}z_{-}$ a rescaling of all the coordinates in AdS can be performed to map it into the problem at hand.

\section{Numerical analysis of the formation of marginally closed trapped surfaces}
\label{sec:cts}

We have seen how the isometries of AdS$_{D}$ can be exploited to rotate a collision of two shock waves with
holographic impact parameter into a collision of two waves sourced at $z=L$ and with spatial impact parameter.
In $D=5$ this latter problem has been solved numerically in \cite{lin_shuryak}. In this section we extend the
numerical solution to other values of the dimension and study the emergence of thresholds for the formation of
the closed trapped surface as a function of the impact parameter.

Before embarking on the search for closed trapped surfaces we have to switch to a 
new system of coordinates $(U,V,Z,\vec{X})$ 
in which the null geodesics are continuous across the location of the waves \cite{gpy1,lin_shuryak}
(see the Appendix for the technical details). 
Once this is done, we look for marginally closed trapped surfaces $\mathcal{S}$ that lie along the position of the 
colliding shock waves $\{U=0,V\leq0\}\cup \{U\leq0,V=0\}$ \cite{flat,gpy1}. This surface can be written as the union of two 
branches $\mathcal{S}=\mathcal{S}_{+}\cup\mathcal{S}_{-}$ parametrized as
\begin{eqnarray}
\mathcal{S}_{+}=\left\{
\begin{array}{l}
U=-\psi_{+}(Z,\vec{X}\,) \\[0.3cm]
V=0
\end{array}
\right., \hspace*{1cm} 
\mathcal{S}_{-}=\left\{
\begin{array}{l}
U=0\\[0.3cm]
V=-\psi_{-}(Z,\vec{X}\,) 
\end{array}
\right. .
\label{ansatz_cts}
\end{eqnarray}
Although most of this surface lies outside of the region $U>0$, $V>0$ where the two waves interact, it ``feels''
the interaction taking place through the matching conditions at the $(D-3)$-dimensional 
intersection of the two branches
$\mathcal{C}=\mathcal{S}_{+}\cap\mathcal{S}_{-}\subset \mathbb{H}_{D-2}$. Imposing that the congruence of 
outer null geodesics normal to $\mathcal{S}$ has zero divergence leads to the equation (see \cite{gpy1} for the
details)
\begin{eqnarray}
\left(\Box_{\mathbb H_{D-2}}-{D-2 \over L^2}\right) \Big(\Phi_{\pm}-\Psi_{\pm}\Big) = 0,
\label{equation_cts}
\end{eqnarray}
where $\Phi_{\pm}(Z,\vec{X}_{\perp})$ are the profiles of the incoming waves and
\begin{eqnarray}
\Psi_{\pm}(Z,\vec{X})={z\over L}\psi_{\pm}(Z,\vec{X}).
\end{eqnarray}
In addition, from the very form \eqref{ansatz_cts}
of the sought marginally closed trapped surface and the requirement that 
the congruence of null geodesics is continuous across the intersection $\mathcal{C}$ we find the conditions
\begin{eqnarray}
\Psi_{\pm}(Z,\vec{X}\,)\Bigg|_{\mathcal{C}}=0, \hspace*{2cm}
g^{ab}\partial_{a}\Psi_{\pm}(Z,\vec{X}\,)\partial_{b}\Psi_{\pm}(Z,\vec{X}\,)\Bigg|_{\mathcal{C}}=4,
\label{matching_cts}
\end{eqnarray}
with $g_{ab}$ the metric of $\mathbb{H}_{D-2}$.

In order to solve the problem \eqref{equation_cts}, \eqref{matching_cts} we solve the differential equation
numerically using the technique employed in Ref. \cite{flatip} for the collision of shock waves in 
flat space-time.  
To implement the procedure we switch to radial coordinates in $\mathbb{H}_{D-2}$ defined in terms of the 
hyperboloid coordinates by
\begin{eqnarray}
Y^{0}=r^{2}+L^{2}, \hspace*{0.5cm} Y^{1}=r\vartheta^{1},\hspace*{1cm}\ldots\hspace*{0.5cm}
Y^{D-2}=r\vartheta^{D-2}
\end{eqnarray}
where $(\vartheta^{1},\ldots,\vartheta^{D-2})$ parametrize the unit sphere $\mathbb{S}^{D-3}$, i.e.,
$(\vartheta^{1})+\ldots+(\vartheta^{D-2})^{2}=1$. Now the metric takes the form
\begin{eqnarray}
ds^{2}_{\mathbb{H}_{D-2}}={dr^{2}\over 1+{r^{2}\over L^{2}}}+r^{2}d\theta^{2}+r^{2}\sin^{2}{\theta} d\Omega^{2}_{D-4}.
\end{eqnarray}
Here we have singled out a polar angle $\theta\in [0,\pi]$ on the sphere $\mathbb{S}^{D-3}$ of constant $r$. 
In these new coordinates\footnote{We have to keep in mind that only $D-4$ of the $D-3$ coordinates $\vartheta^{i}$ parametrizing
the sphere $\mathbb{S}^{D-4}$ are independent, since $(\vartheta^{1})+\ldots+(\vartheta^{D-3})^{2}=1$.} 
$(r,\theta,\vartheta^{i})$, with $i=1,\ldots,D-3$,
the sources of the incoming waves are located respectively at the points $({b\over 2},0,0,\ldots,0)$ and
$({b\over 2},\pi,0,\ldots,0)$. Hence, the residual O($D-3$) symmetry is generated by the isometries of the 
sphere $\mathbb{S}^{D-4}$ at constant $r$ and constant 
$\theta$. As a consequence of this symmetry, both the profiles of
the incoming waves $\Phi_{\pm}(r,\theta)$ and the functions defining the two branches of the trapped surface
$\Psi_{\pm}(r,\theta)$ only depend on $r$ and 
$\theta$. In addition, the surface $\mathcal{C}$ is generated by the action of O($D-3$) 
on a curve that we parametrize as $r=LG(\theta)$ for $0\leq\theta\leq\pi$. We have introduced
the AdS radius $L$ in order to make the function $G(\theta)$ dimensionless.

Introducing the function $H_{\pm}(r,\theta)\equiv \Phi_{\pm}(r,\theta)
-\Psi_{\pm}(r,\theta)$, Eq. \eqref{equation_cts} reads in these coordinates
\begin{eqnarray}
\left[\left(1+{r^{2}\over L^{2}}\right)\partial_{r}^{2}+{(D-3)L^{2}+(D-2)r^{2}
\over r L^{2}}\partial_{r}+{1\over r^{2}}\partial_{\theta}^{2}+{D-4\over
r^{2}\tan{\theta}}\partial_{\theta}-{D-2\over L^{2}}\right]H_{\pm}
=0
\label{equation2}
\end{eqnarray}
with the boundary condition 
\begin{eqnarray}
H_{\pm}(r,\theta)\Bigg|_{r=LG(\theta)}\hspace*{-0.4cm}=\Phi_{\pm}\Big(LG(\theta),\theta\Big).
\end{eqnarray}
This problem has to be solved in the region $0\leq r\leq LG(\theta)$, $0\leq\theta\leq\pi$, i.e. those points in the
upper half-plane limited by the curve $r=LG(\theta)$.
Because the sources of the waves are located symmetrically around $r=0$ the function $G(\theta)$ has the symmetry
\begin{eqnarray}
G(\theta)=G(\pi-\theta),
\end{eqnarray}
whereas $\Psi_{\pm}(r,\theta)$ and $\Phi_{\pm}(r,\theta)$ satisfy
\begin{eqnarray}
\Psi_{\pm}(r,\theta)=\Psi_{\pm}(r,\pi-\theta), \hspace*{1cm} 
\Phi_{\pm}(r,\theta)=\Phi_{\mp}(r,\pi-\theta).
\label{symmetry_conditions}
\end{eqnarray}
The boundary $G(\theta)$ is determined by imposing that the solution to the boundary problem satisfies 
the addition constraint
\begin{eqnarray}
\left[\left(1+{r^{2}\over L^{2}}\right)(\partial_{r}\Psi_{+})(\partial_{r}\Psi_{-})+{1\over r^{2}}(\partial_{\theta}\Psi_{+})
(\partial_{\theta}\Psi_{-})
\right]\Bigg|_{r=LG(\theta)}\hspace*{-0.4cm}=4.
\label{addition_constraint}
\end{eqnarray}

In the symmetric collision that we are going to consider in the following $\Phi_{+}(z,\vec{x})=\Phi_{-}(z,\vec{x})$
and as a consequence the trapped surface has to be symmetric under a parity transformation in the longitudinal 
coordinate. This means that $H_{+}(r,\theta)=H_{-}(r,\theta)\equiv H(r,\theta)$, and we have that 
the symmetry conditions 
\eqref{symmetry_conditions} imply that
\begin{eqnarray}
\partial_{\theta}H(r,0)=\partial_{\theta}H(r,\pi)=0.
\end{eqnarray}

The numerical solution of the problem at hand is complicated by the fact that the function $G(\theta)$
to be determined appears in the boundary conditions of the differential equation. Because of this we would like
to make a change of coordinates that transforms the complicated integration region into a simpler one. 
Following \cite{flatip}, we define the new dimensionless radial coordinate $\rho$ by
\begin{eqnarray}
r=LG(\theta)\rho.
\end{eqnarray}
Now, in terms of this new variable Eq. \eqref{equation2} reads
\begin{eqnarray}
\left\{\left[1+{G^{2}\rho^{2}}+\left({G'\over G}\right)^{2}\right]\partial_{\rho}^{2}+{1\over \rho^{2}}
\partial_{\theta}^{2}-{2\over \rho}{G'\over G}\partial_{\rho}\partial_{\theta}+{1\over \rho}
\left[
2\left({G'\over G}\right)^{2}-{G''\over G}-{D-4\over \tan{\theta}}\left({G'\over G}\right)
\right.\right. \nonumber \\
\label{boundary_problem_final}\\[-0.3cm]
+\left.\left.{(D-3)+(D-2)G^{2}\rho^{2}}\right]\partial_{\rho}
+{D-4\over\rho^{2}\tan{\theta}}
\partial_{\theta}
-(D-2)G^{2}\right\}H(\rho,\theta)
=0. \nonumber
\end{eqnarray}
With this we are left with a boundary problem in the half-circle  
$\rho=1$ with $0\leq\theta\leq\pi$, whereas the function $G(\theta)$ defining the boundary in the original coordinates
$(r,\theta)$ appears now in the equation. The boundary conditions on the lower and upper boundaries of the 
upper half-circle are respectively
\begin{eqnarray}
\partial_{\theta}H(\rho,0)=\partial_{\theta}H(\rho,\pi)=0, \hspace*{1cm} \partial_{\theta}H(\rho=1,\theta)=
{d\over d\theta}\Phi\Big(r=LG(\theta),\theta\Big).
\label{boundary_cond_final}
\end{eqnarray}
This means that we have von Neumann boundary conditions on the lower boundary and Dirichlet condition on 
the upper one. To 
derive the last equation in \eqref{boundary_cond_final} 
we have used that the function defining the trapped surface vanishes on $\mathcal{C}$, $\Psi(\rho=1,\theta)=0$, 
and as a consequence $\partial_{\theta}\Psi(\rho=1,\theta)=0$.
This equation actually simplifies the additional condition \eqref{addition_constraint}
\begin{eqnarray}
\left[1+{G'(\theta)^{2}\over G(\theta)^{2}}+{G(\theta)^{2}}\right]\Big[\partial_{\rho}
\widetilde{\Psi}(1,\theta)\Big]^{2}
=4G(\theta)^{2}.
\label{additional_cond_final}
\end{eqnarray}
In order to implement the numerical algorithm it is convenient to introduce the dimensionless functions
$\widetilde{\Psi}(\rho,\theta)\equiv {1\over L}\Psi(\rho,\theta)$, $\widetilde{\Phi}(\rho,\theta)
\equiv {1\over L}\Phi(\rho,\theta)$ and $\widetilde{H}(\rho,\theta)\equiv {1\over L}H(\rho,\theta)$.

\begin{figure}[t]
\centerline{
\includegraphics[width=4.9cm]{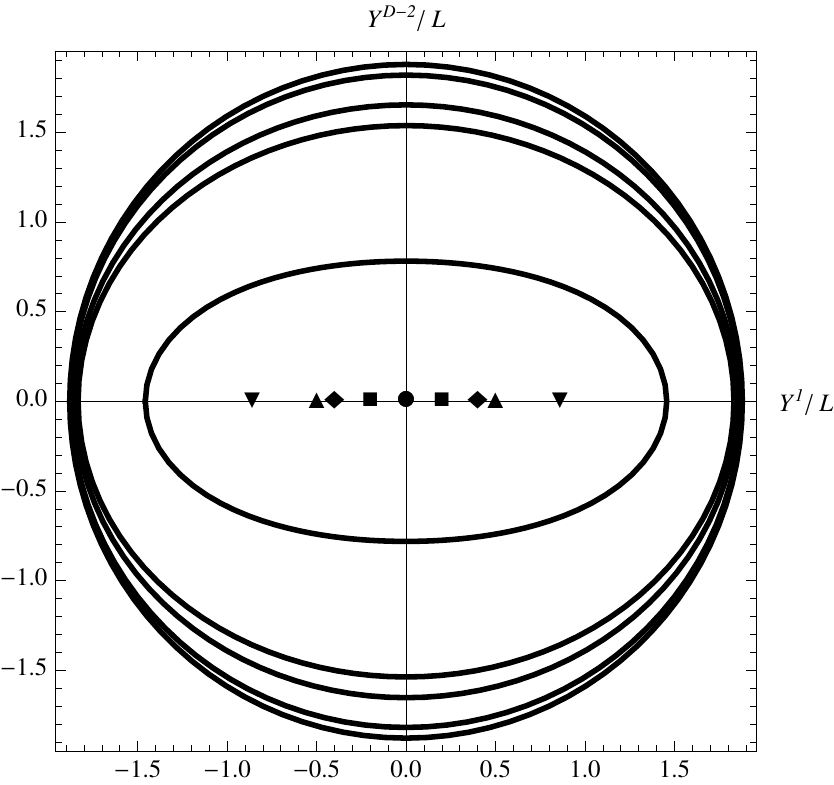}
\hspace*{0.75cm}
\includegraphics[width=5.0cm]{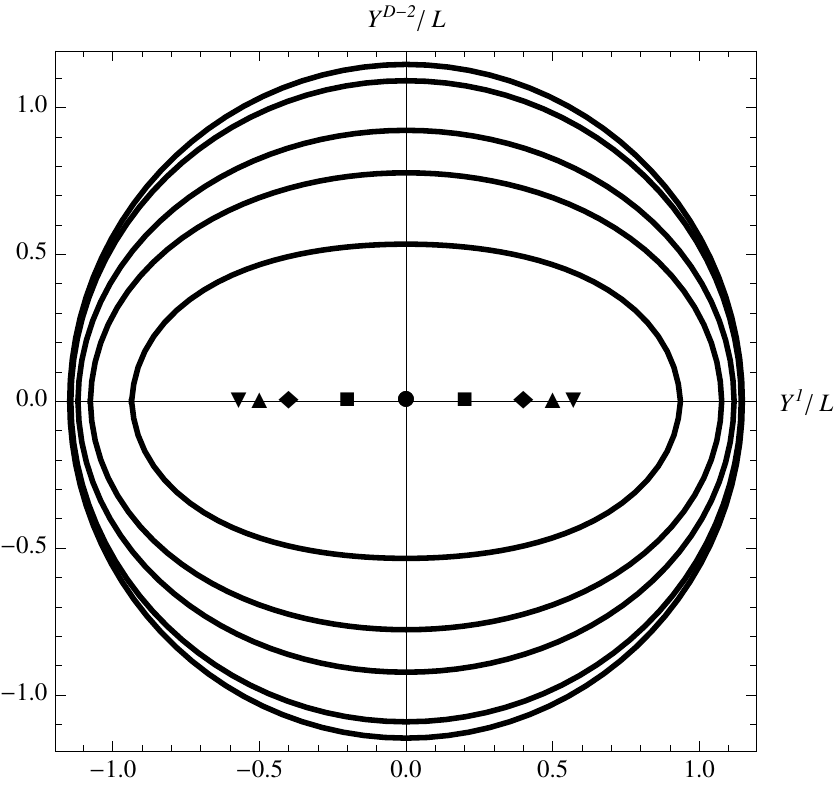}
\hspace*{0.75cm}
\includegraphics[width=5.0cm]{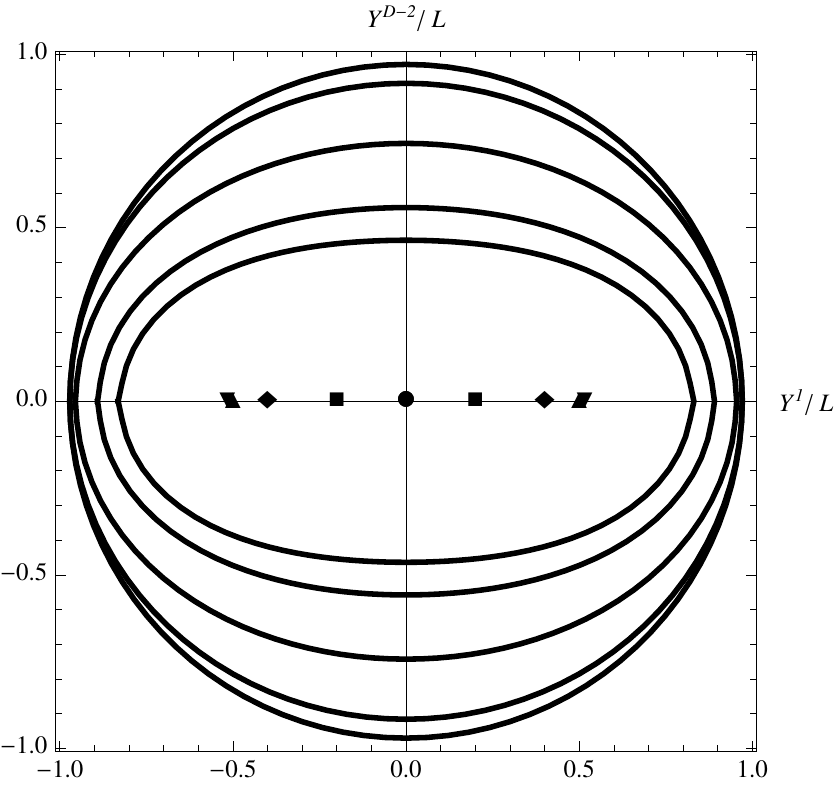}
}

\centerline{
\includegraphics[width=5.0cm]{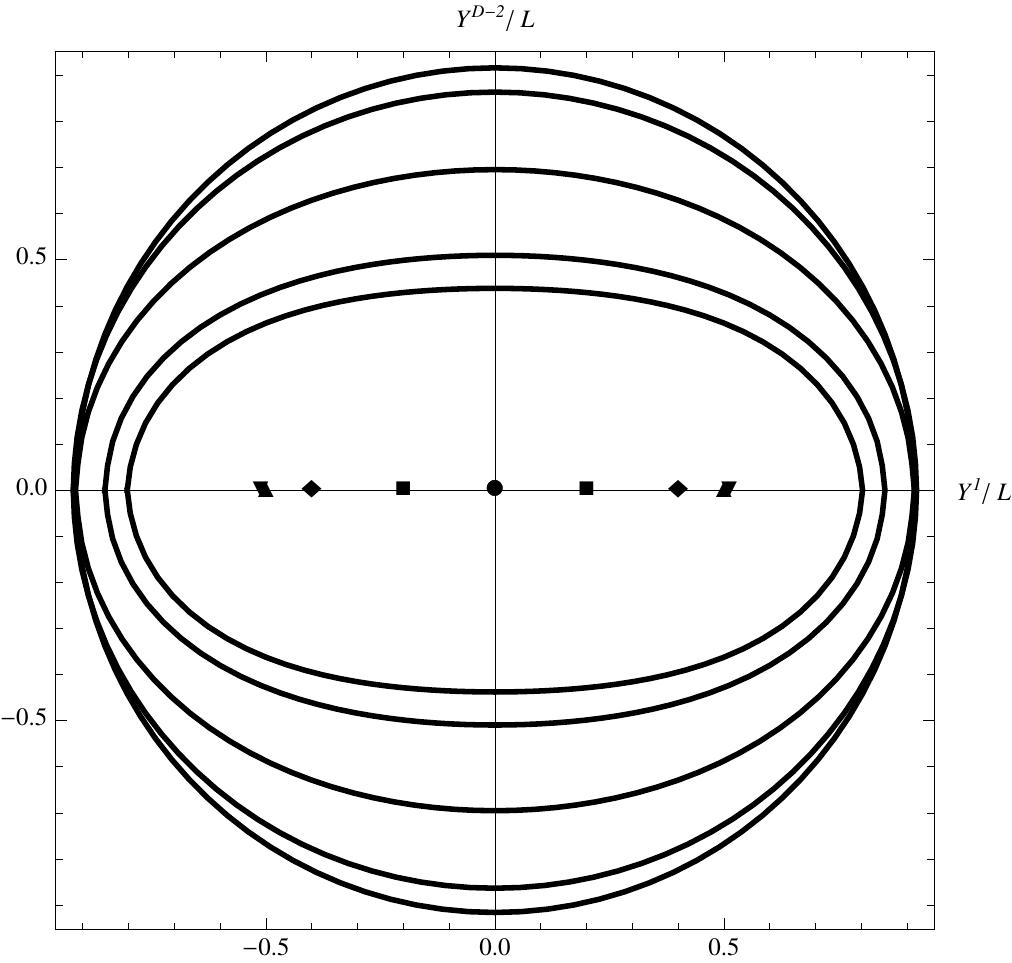}
\hspace*{0.75cm}
\includegraphics[width=5.0cm]{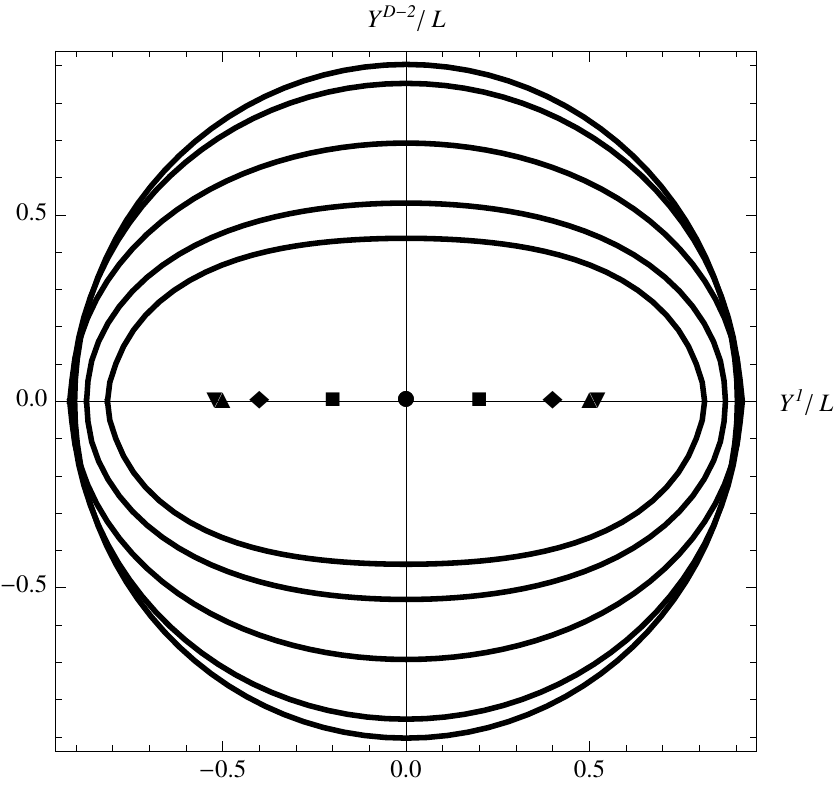}}
\caption{Numerical solutions for the shape of the section of the surface $\mathcal{C}$ 
on the plane $Y^{1}$-$Y^{D-2}$ in hyperbolic coordinates for $D=4,5,6,7$ and 8 for various values of the 
impact parameter $b$ and $\mu={L^{D-3}\over G_{N}}$. 
From the outer inwards the curves correspond to ${b\over L}=0$ ({\large $\bullet$}), 
${b\over L}=0.2$ ({\scriptsize${\blacksquare}$}), ${b\over L}=0.4$ ({\footnotesize $\blacklozenge$}), 
${b\over L}=0.5$ ({\footnotesize $\blacktriangle$}) and
the critical value ${b_{c}\over L}$ ({\footnotesize $\blacktriangledown$}). The symbols indicate the position of the
incoming sources.
}
\label{fig2}
\end{figure}

We can now attack the numerical solution of the boundary problem \eqref{boundary_problem_final}, 
\eqref{boundary_cond_final} subjected to the additional condition \eqref{additional_cond_final}. 
We follow the procedure used by the authors of \cite{flatip} and choose a trial function $G_{1}(\theta)$ to solve the boundary problem \eqref{boundary_problem_final}, \eqref{boundary_cond_final} using a method
of finite differences. This gives a solution  
$\widetilde{H}_{1}(\rho,\theta)$ 
in terms of which we compute $\widetilde{\Psi}_{1}(\rho,\theta)=\widetilde{\Phi}(\rho,\theta)-
\widetilde{H}_{1}(\rho,\theta)$.
In the extremely unlikely case in which $G_{1}(\theta)$ were the solution of the problem the function
\begin{eqnarray}
T_{1}(\theta)\equiv \left[1+\left({G'\over G}\right)^{2}+{G^{2}}\right]
\Big[\partial_{\rho}\widetilde{\Psi}_{1}(1,\theta)\Big]^{2}-4G^{2}
\label{t1function}
\end{eqnarray}
would vanish. However, since this is not going to be the case we introduce a new trial function 
\begin{eqnarray}
G_{2}(\theta)=G_{1}(\theta)+\epsilon\,T_{1}(\theta),
\end{eqnarray}
with $\epsilon>0$ a small number, and solve again the boundary problem \eqref{boundary_problem_final},
\eqref{boundary_cond_final} obtaining a second function $\widetilde{H}_{2}(\rho,\theta)$ and 
$\widetilde{\Psi}_{2}(\rho,\theta)$. 
Computing now $T_{2}(\theta)$ as in Eq. \eqref{t1function} and 
iterating the algorithm a sufficient number of times a numerical solution to 
the problem can be found with the required precision.

We have used the procedure described above to find the solutions to the trapped surface equation in AdS$_{D}$ with $D=4$, 5, 6, 7 and 8. 
We have solved the boundary problem using a finite difference method with a grid of 50 points in the angular
coordinate and 100 points in the radial one. With $\epsilon=10^{-4}$ we find good convergence to the solution. 
In order to handle the coordinate singularity at $r=0$ we have imposed the von Neumann boundary
condition $\partial_{n}H=0$ at this point. 

Our numerical analysis shows the existence in all cases of a critical value of the impact parameter 
above which no solution to the marginally closed trapped surface equation is found. This is qualitatively similar to 
the corresponding situation in flat space \cite{flatip} as well as in AdS$_{5}$ \cite{lin_shuryak}. In 
Fig. \ref{fig2} we plot the section of the trapped surface on the $Y^{1}$-$Y^{D-2}$ plane for
a symmetric collision in $D=4,5,6,7$ and
$8$ with energy $\mu=G_{N}^{-1}L^{D-3}$. 
For AdS$_{5}$ our numerical solutions 
are in perfect agreement with the results found by the authors of \cite{lin_shuryak} using a different numerical method.

\begin{figure}[t]
\centerline{
\includegraphics[width=8cm]{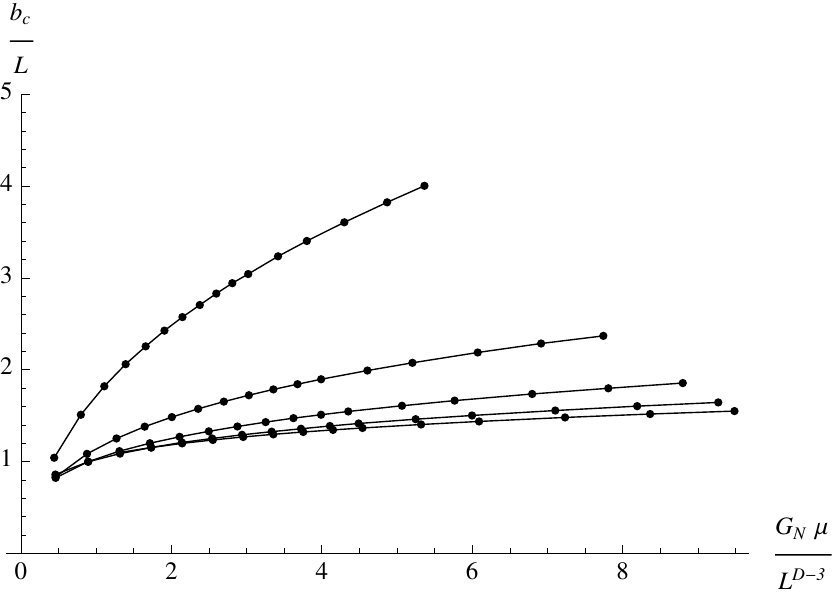}\hspace*{1cm}
\includegraphics[width=8cm]{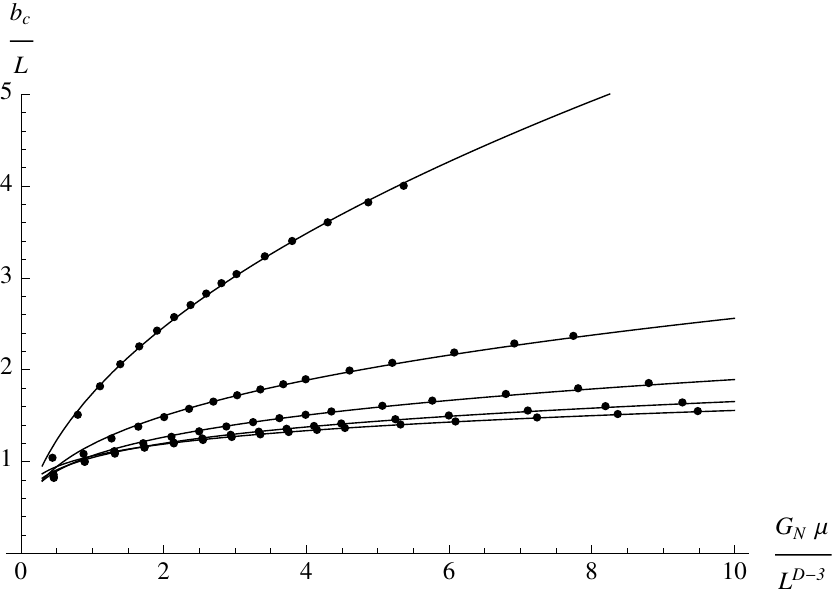}}
\caption{\footnotesize{In the left panel we plot the critical impact parameter ${b_c\over L}\equiv{b_{+}-b_{-}\over L}$ as a function of  
${G_N \mu\over L^{D-3}}$ for the symmetric collision of two 
shock waves with sources located at $z=L$. The curves correspond, from top to bottom, to $D=4,5,6,7$ and $8$. On the right
panel the same points are shown together with a fit to a function with the scaling of Eq. \eqref{scaling_b_E}.}}
\label{figure4}
\end{figure}

On the left panel of Fig. \ref{figure4}, on the other hand, the values of the critical impact parameter
as a function of the energy are shown for different dimensions. As can be seen in the right panel of Fig. 
\ref{figure4}, the results are very well fitted by 
the following scaling of the critical impact parameter with the energy of the incoming waves 
\begin{eqnarray}
{b_{c}\over L}\sim \left({G_{N}\mu\over L^{D-3}}\right)^{1\over D-2},
\label{scaling_b_E}
\end{eqnarray}
where the proportionality constant is of order one.
One consequence of this scaling is that the dependence of the critical impact parameter with the energy, 
measured in units of the AdS$_{D}$ energy $L^{D-3}G_{N}^{-1}$, 
flattens as the dimension increases. In the limit $D\rightarrow\infty$ (with 
$L^{D-3}G_{N}^{-1}$ fixed) this seems to implicate that the critical impact parameter is constant with the energy.

In all our previous analysis we have focused our attention on 
physical situations where the energy of the incoming waves are larger or of the order
of the AdS energy scale. This means that the physics is sensitive to the large-scale geometry of AdS. 
Flat space is retrieved by taking
$L\rightarrow\infty$ with the energy and impact parameter of the collision fixed. Since in this limit the energy 
is very small compared with the AdS natural energy scale, the flat space behavior of the critical impact parameter with the
energy has to be recovered in the plots of Fig. \ref{figure4} in the region around the origin.
On dimensional grounds, the scaling between the critical impact parameter and the energy for gravitational shock waves collision 
in flat space has to be of 
the form $b_{c}\sim (G_{N}\mu)^{1\over D-3}$. A crossover between 
\eqref{scaling_b_E} and 
this scaling should take place as the energy of the collision is much smaller than the AdS energy scale.

To compare our results with the analytical study of the formation of trapped surfaces in the collision of shock waves in 
AdS$_{5}$ with nonvanishing impact parameter carried out in \cite{gpy2},
we notice that the authors of this reference work in the limit where the energy of the collision
satisfies 
\begin{eqnarray}
\left({G_{N}\mu\over L^{D-3}}\right)^{1\over D-2}\gg 1,
\end{eqnarray} 
while keeping fixed the size of the impact parameter. Because of the scaling \eqref{scaling_b_E}
the results of \cite{gpy2} are in the regime where the impact parameter is always much smaller than the 
critical value and, as a consequence, a marginally closed trapped surface always forms.

\section{Discussion}
\label{sec:concl}

\begin{figure} 
\centerline{\includegraphics[width=7.5cm]{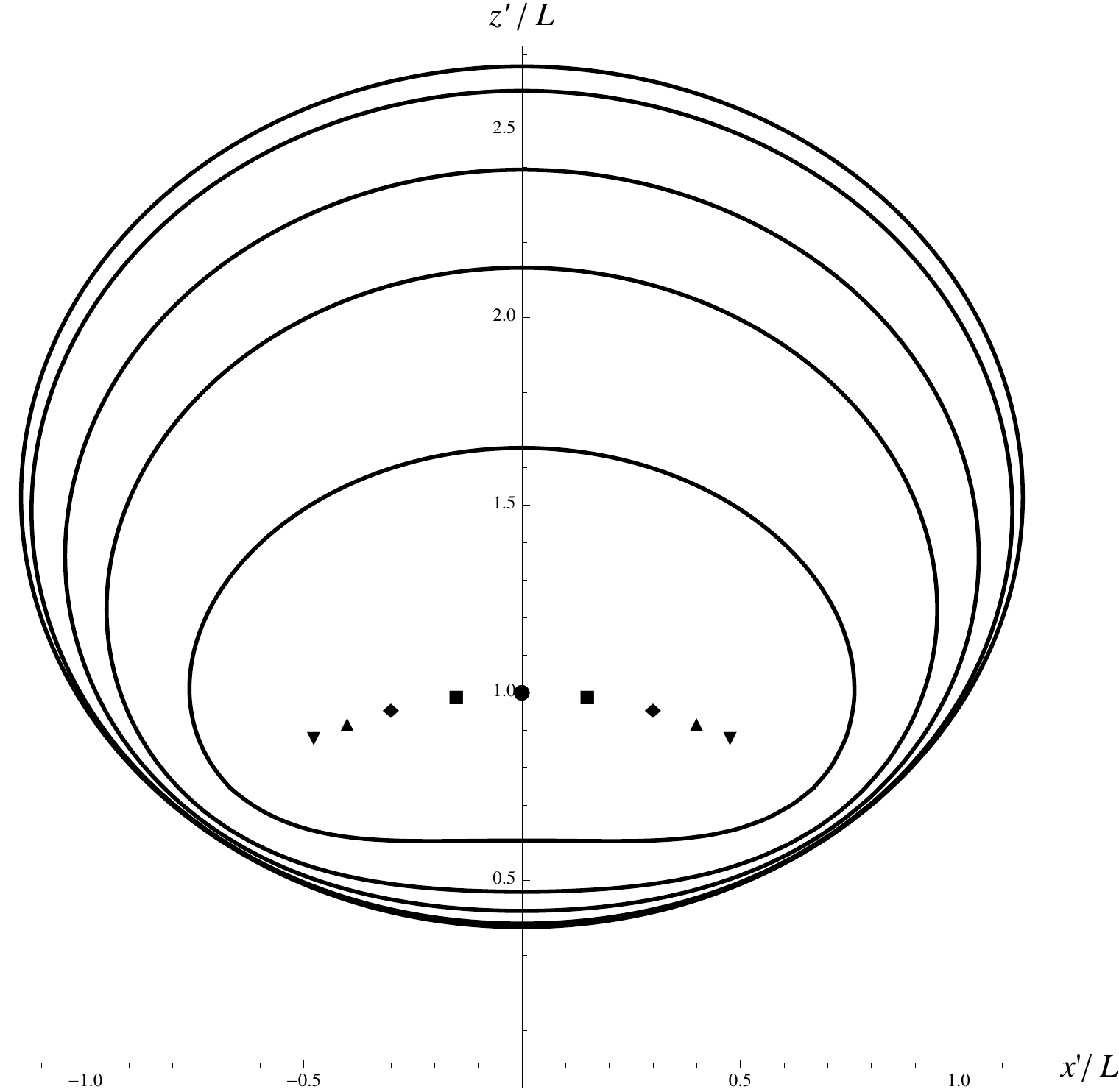}
\hspace*{1cm}\includegraphics[width=7.5cm]{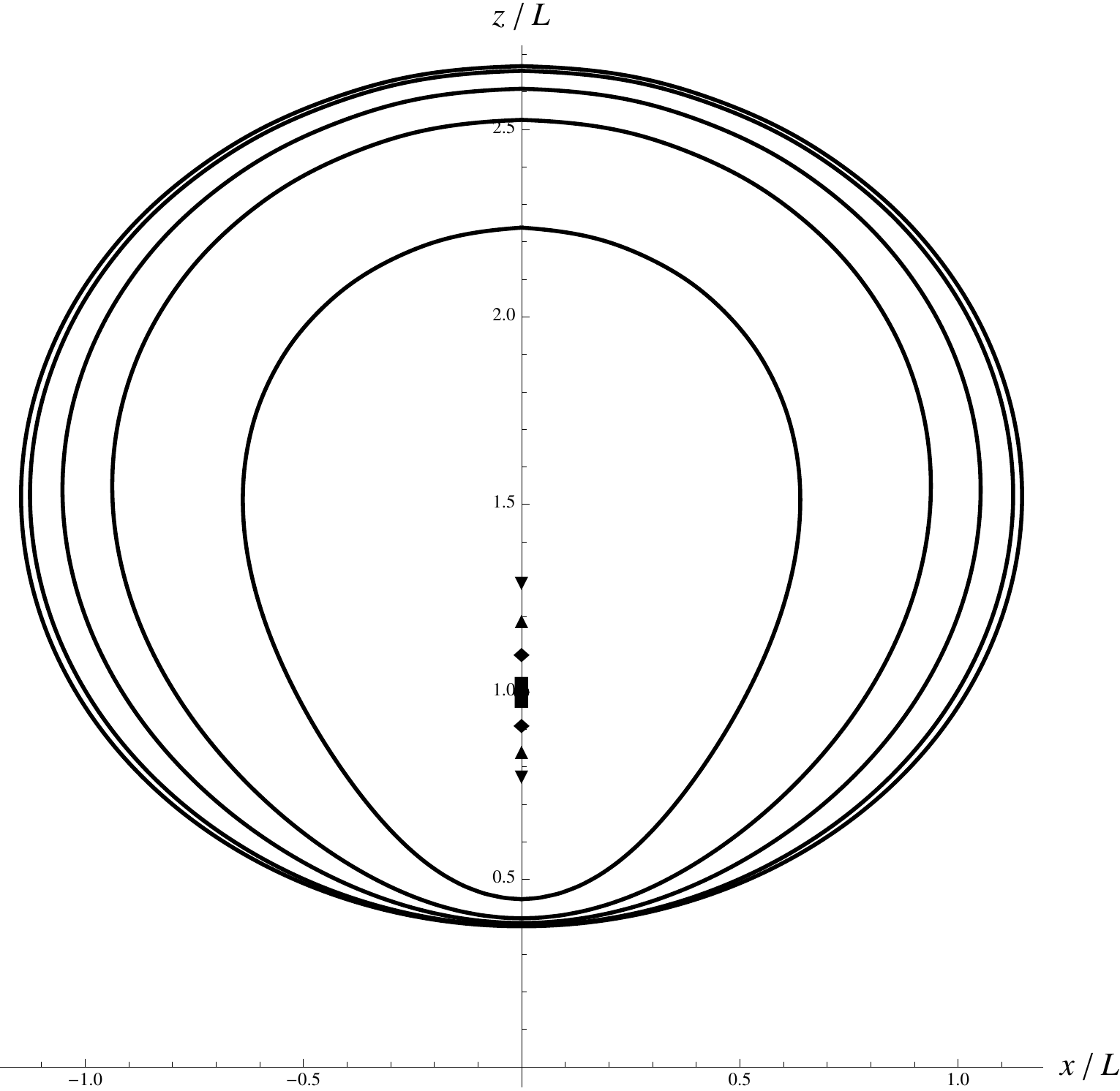}}
\caption{\footnotesize{In the left panel we plot the 
section of the closed trapped surfaces in Poincar\'e coordinates $(x',z')$
shown in Eqs. \eqref{fullrotations} and \eqref{coordinates''}. The curves, from outer inwards, 
correspond to the following
values of the impact parameter: ${b\over L}=0$ ({\large $\bullet$}), 
${b\over L}=0.3$ ({\scriptsize${\blacksquare}$}), ${b\over L}=0.6$ ({\footnotesize $\blacklozenge$}), 
${b\over L}=0.8$ ({\footnotesize $\blacktriangle$}) and
the critical value ${b_{c}\over L}$ ({\footnotesize $\blacktriangledown$}). On the right panel we see
the sections of the 
closed trapped surfaces after inverting the rotations \eqref{fullrotations} back to the coordinates $(x,z)$. In both 
plots the horizontal axis representes the boundary of $\mathbb{H}_{3}$.}}
\label{fig5}
\end{figure}

Having obtained the numerical solution for the collision of two shock waves with ``spatial'' impact parameter we can apply the 
transformations described in the previous section to get solutions of the closed trapped surface equation for the collision of two
waves with ``holographic'' impact parameter. On the left panel of Fig. \ref{fig5} we have plotted, for various values of the
spatial impact parameter, the section of the marginally 
trapped surface for a symmetric collision in AdS$_{5}$ of two waves with energies $\mu_{\pm}=G_{N}L^{-2}$. Unlike the plots shown in Fig.
\ref{fig2} here we have used
the Poincar\'e coordinates $(x',z')$ defined in Eq. \eqref{sources_step1}. Hence, in this plot the boundary of AdS$_{5}$ is 
located along the horizontal axis. 
On the right panel of Fig. \ref{fig5}, on the other hand, we see the section of the trapped surfaces
previous to the O(2) rotation of angle ${\pi\over 2}$ described in Section \ref{sec:gw}. Now the sources lie in the $z$ axis and therefore
the contours show the section of the trapped surface formed as a result of a collision of two waves with holographic impact parameter.
The innermost curve represents the surface corresponding to the critical value of the impact parameter, above which no 
marginally closed trapped surface of the type studied here exits.

The conclusion is the existence of a maximum value of the holographic impact parameter for the formation 
of this class of trapped surfaces. Taking this fact at face value, one 
could 
be tempted to interpret it holographically 
as signaling a threshold for the formation of a thermal CFT plasma. Thus, thermalization would not take place when 
the size difference between the two
colliding energy lumps is larger than the critical value. This would actually make sense on physical 
grounds: if one of the colliding lumps is much smaller than the other it might not have enough degrees of freedom to 
induce a thermal state in the whole system as a result of the collision. One should be reminded, however, that the 
non existence of a closed trapped surface of the type studied above does not rule out the possibility of other trapped surfaces
being formed in the process. Therefore, in order to decide whether a threshold for thermalization exists as
a function of the difference in size of the colliding objects one would need
to solve for the geometry of the interaction region of Fig. \ref{fig:1}. This problem, together with other issues such as the
thermalization time for subcritical collisions, will be addressed elsewhere. 

In the dual conformal field theory, the O(2) transformation that maps a collision of two equal objects with impact parameter into
the head-on collision of two lumps of different size is a combination of a translation and a special conformal transformation 
in one of the transverse coordinates, inherited from the AdS$_{D}$ 
bulk transformation \eqref{finite_transf_bulk}. Interestingly, this relates a physical
process with nonvanishing angular momentum to another one where the angular momentum is zero. In the gravitational dual, on the
other hand, the combination [see Eqs. \eqref{sources2} and \eqref{energy_after_rotation}]
\begin{eqnarray}
\mathcal{Q}_{\pm}=\mu_{\pm}'(\theta)L\left[1+{z_{\pm}'(\theta)^{2}+{b}_{\pm}'(\theta)^{\,2}\over L^{2}}\right]
\label{invariantQpm}
\end{eqnarray}
is independent of the angle $\theta$ and therefore invariant under the O(2) rotation. Holographically, 
this relates the impact parameter, the size of the 
lumps and their energies. A second invariant can be also constructed as 
\begin{eqnarray}
\mathcal{Q}(\theta)=\mu_{+}'(\theta)\mu_{-}'(\theta)z_{+}'(\theta)z_{-}'(\theta)
\label{invariantQ}
\end{eqnarray}
that mixes the energies of the colliding objects with their sizes.

In $D=5$ the collision of two gauge theory energy 
lumps with large size difference could be used as a model for hadron-nucleus collisions at strong coupling. 
Here we have considered the case of 
head-on scattering. Collisions of unequal objects with a nonvanishing impact parameters can be studied as well by performing O(2) 
rotations on the numerical solutions obtained in the previous section with an angle $|\theta|<{\pi\over 2}$. It would be 
interesting to see if the invariants \eqref{invariantQpm} and \eqref{invariantQ} have any relevance in the 
phenomenological description of this type of collisions.

\section*{Acknowledgments}

We thank Luis \'Alvarez-Gaum\'e, Roberto Emparan, Kerstin E. Kunze and Agust\'{\i}n Sabio Vera for discussions. 
A.D.-V. acknowledges support from a
Castilla y Le\'on Regional Government 
predoctoral fellowship and Spanish Government Grant FIS2009-07238. M.A.V.-M. has been partially
supported by Spanish Government Grants FPA2009-10612 and FIS2009-07238, 
Basque Government Grant IT-357-07 and Spanish Consolider-Ingenio 2010 Programme CPAN (CSD2007-00042).

\section*{Appendix}
\renewcommand{\thesection}{A}

In this Appendix we give some details on the change of coordinates that have to be used to 
eliminate the delta-function terms in the metric of the shock wave and therefore to make the congruence
of null geodesics continuous across the wave fronts $u=0$, $v=0$. 
Let us begin with the metric \eqref{metric}
now written in the form
\begin{eqnarray}
ds^{2}_{+}={L^{2}\over z^{2}}\Big[dz^{2}-dudv+d\vec{x}^{\,2}+\varphi_{+}(z,\vec{x})
\delta(u)du^{2}\Big].
\label{metric_app}
\end{eqnarray}
and introduce the new coordinates $(U,V,Z,\vec{X})$ defined by
\begin{eqnarray}
u&=& U, \nonumber \\
v&=& V+\varphi_{+}(Z,\vec{X})\theta(U)+{1\over 4}U\theta(U)
\Big[\vec{\nabla}\varphi_{+}(Z,\vec{X})\Big]^{2}
+{1\over 4}U\theta(U)\left[\partial_{Z}\varphi_{+}(Z,\vec{X})\right]^{2},
\nonumber \\
z&=& Z+{1\over 2}U\theta(U)\partial_{Z}\varphi_{+}(Z,\vec{X}),
\label{newcoordinatesAdS}\\
\vec{x} &=& \vec{X}+{1\over 2}U\theta(U)\vec{\nabla}\varphi_{+}(Z,\vec{X}).
\nonumber
\end{eqnarray}
The presence of the step function $\theta(U)$ in the change of coordinates is necessary in order
to eliminate the delta-function terms in the metric, responsible for the discontinuity
of the geodesics. After a long but simple calculation we find 
that the metric \eqref{metric_app} for the first wave takes the form
\begin{eqnarray}
ds^{2}_{+}={-dUdV+\mathcal{H}_{ab}^{(+)}\mathcal{H}_{bc}^{(+)}dX^{a}dX^{c}\over
\Big[Z+{1\over 2}U\theta(U)\partial_{Z}\varphi_{+}\Big]^{2}},
\label{metricnewcoordAdSU}
\end{eqnarray}
with
\begin{eqnarray}
\mathcal{H}^{(+)}_{ab}=\delta_{ab}
+{1\over 2}U\theta(U)\partial_{a}\partial_{b}\varphi_{+}(Z,\vec{X}).
\end{eqnarray}
The indices $a,b,c$ run over all the spatial coordinates $(Z,\vec{X})$. The line element
\eqref{metricnewcoordAdSU} is the AdS analog of the 
so-called Rosen form of a plane wave in flat space-time, whereas the metric \eqref{metric_app}
we started with is the Brinkmann form of the wave.

The analysis can be repeated for the second wave, whose metric can be obtained from Eq.
\eqref{metric_app} by changing $u\rightarrow v$ and $\varphi_{+}(z,\vec{x})
\rightarrow \varphi_{-}(z,\vec{x})$. Implementing then the change of 
coordinates \eqref{newcoordinatesAdS}
with the replacements $U\rightarrow V$, $u\rightarrow v$ and $\varphi_{+}(Z,\vec{X})\rightarrow
\varphi_{-}(Z,\vec{X})$ we arrive at the metric
\begin{eqnarray}
ds^{2}_{-}={-dUdV+\mathcal{H}_{ab}^{(-)}\mathcal{H}_{bc}^{(-)}dX^{a}dX^{c}\over
\Big[Z+{1\over 2}V\theta(V)\partial_{Z}\varphi_{2}\Big]^{2}},
\label{metricnewcoordAdSV}
\end{eqnarray}
where now the metric function $\mathcal{H}^{(-)}_{ab}$ is given by
\begin{eqnarray}
\mathcal{H}^{(-)}_{ab}=\delta_{ab}+{1\over 2}V\theta(V)\partial_{a}\partial_{b}\varphi_{2}(Z,\vec{X}).
\end{eqnarray}
It is important to notice that the metric elements \eqref{metricnewcoordAdSU} and \eqref{metricnewcoordAdSV}
coincide in the lower wedge $U<0$, $V<0$ where the line element of AdS$_{D}$ is recovered. 
Hence, the metric in the whole region outside the interaction 
wedge $U>0$, $V>0$ can be written as
\begin{eqnarray}
ds^{2}={-dUdV+\Big[\mathcal{H}_{ab}^{(+)}\mathcal{H}_{bc}^{(+)}+
\mathcal{H}_{ab}^{(-)}\mathcal{H}_{bc}^{(-)}-\delta_{ab}\Big]dX^{a}dX^{c}\over
\Big[Z+{1\over 2}U\theta(U)\partial_{Z}\varphi_{+}+
{1\over 2}V\theta(V)\partial_{Z}\varphi_{-}\Big]^{2}}.
\end{eqnarray}

The next step is to write the Einstein equations in the new coordinates. This task, however, is simplified
by noticing that for the line element of the first wave \eqref{metric_app} the only nonvanishing component of 
both $G_{\mu\nu}+\Lambda g_{\mu\nu}$ and $T_{\mu\nu}$ is the $u$-$u$ component. Since $U=u$ we find again that 
the equation for $\varphi_{+}(Z,\vec{X})$ is given by $G_{UU}+\Lambda g_{UU}=8\pi G_{N}T_{UU}$ and after
a bit of algebra Eq. \eqref{ee}
is recovered in the new coordinates (the same argument applies to the second wave with the replacement
$U\rightarrow V$) 
\begin{eqnarray}
\left(\Box_{\mathbb{H}_{D-2}}-{D-2\over L^{2}}\right)\Phi_{\pm}(Z,\vec{X})=
-16\pi G_{N}\mu_{\pm}\left({Z_{\pm}\over L}\right)^{D-1}\delta(Z-z_{\pm})\delta^{(D-3)}(\vec{X}-\vec{b}_{\pm}),
\end{eqnarray}
where $\Phi_{\pm}(Z,\vec{X})={L\over Z}\varphi_{\pm}(Z,\vec{X})$.
In fact, the reason behind obtaining the same equations in both coordinate systems is that the change 
$(z,\vec{x})\rightarrow (Z,\vec{X})$ is trivial in the transverse space $\mathbb{H}_{D-2}$. This can seen inmediately
by setting $U=0$ in \eqref{newcoordinatesAdS}, and respectively 
$V=0$ in the corresponding change of coordinates for the second
wave.

\end{document}